\documentclass[10pt, conference, compsocconf]{IEEEtran}
\usepackage{cite}
\usepackage{graphicx}
\usepackage{subfigure}
\graphicspath{ {images/} }
\usepackage{url}
\usepackage{array}
\usepackage{multirow}
\usepackage{xcolor}
\usepackage[utf8]{inputenc}
\usepackage{etoolbox}
\makeatletter
\patchcmd{\@makecaption}
  {\scshape}
  {}
  {}
  {}
\makeatother

\usepackage[nameinlink]{cleveref}
\crefname{subsection}{\S}{\S}
\usepackage{listings}

\def\etal{{\it et al.}\hspace{0.1pc}}
\def\etc{{\it etc.}\hspace{0.1pc}}

\begin{document}

\title{Low Overhead Instruction Latency Characterization for NVIDIA GPGPUs}

\author
{
\IEEEauthorblockN
{
Yehia Arafa$^1$, Abdel-Hameed A. Badawy$^{1,2}$, Gopinath Chennupati$^2$, Nandakishore Santhi$^2$, and Stephan Eidenbenz$^2$ 
}

\IEEEauthorblockA
{
$^1$ Klipsch School of ECE, New Mexico State University. Las Cruces, NM, USA\\
\{yarafa, badawy\}@nmsu.edu\\
$^2$ Los Alamos National Laboratory. Los Alamos, NM, USA\\
   \{gchennupati, nsanthi, eidenben\}@lanl.gov\\
}
\vspace{-21pt}
}

\maketitle

\begin{abstract}
The last decade has seen a shift in the computer systems industry where heterogeneous computing has become prevalent. Graphics Processing Units (GPUs) are now present in supercomputers to mobile phones and tablets. GPUs are used for graphics operations as well as general-purpose computing  (GPGPUs) to boost the performance of compute-intensive applications. However, the percentage of undisclosed characteristics beyond what vendors provide is not small. In this paper, we introduce a very low overhead and portable analysis for exposing the latency of each instruction executing in the GPU pipeline(s) and the access overhead of the various memory hierarchies found in GPUs at the micro-architecture level. Furthermore, we show the impact of the various optimizations the CUDA compiler can perform over the various latencies. We perform our evaluation on seven different high-end NVIDIA GPUs from five different generations/architectures: {\em Kepler}, {\em Maxwell}, {\em Pascal}, {\em Volta}, and {\em Turing}. The results in this paper can help architects to have an accurate characterization of the latencies of these GPUs, which will help in modeling the hardware accurately. Also, software developers can perform informed optimizations to their applications.
\end{abstract}

\begin {IEEEkeywords}
GPGPUs, Latency, PTX, Benchmarking, High-Level Optimizations, Turing, CUDA
\end{IEEEkeywords}

\section{Introduction}
\label{sec:intro}
Graphics Processing Units (GPUs) were originally designed to accelerate graphics operations. Now, they have become one of the most crucial hardware components of computing systems. Over the last decade, GPUs have evolved to be powerful co-processors that perform general non-specialized calculations that would typically be performed by the CPU. General Purpose Graphic Processing Units (GPGPUs) are now fundamental components in any high-performance computing (HPC) system due to their ability to perform complex computations efficiently. The emerging of AI, machine learning, deep learning, bitcoin mining has pushed GPGPUs over the top in popularity and versatility way beyond gaming. According to the recent rank of the top 500 most powerful distributed computer systems in the world (TOP500 list)~\cite{top500}, systems that consist of GPUs such as Summit, 95$\%$ of its peak performance (187.7 petaflops) is derived from 27,686 GPUs. This is mainly due to the high computational power the recent GPUs have. For instance, the NVIDIA Tesla V100 GPU is capable of deriving peak computational rates of 7.8 TFLOPS for double-precision floating-point (FP64) performance and 15.7 TFLOPS for single-precision (FP32) performance.

Over the last decade, NVIDIA has introduced seven different GPU generations/architectures~\cite{tesla,fermi,kepler,maxwell,pascal, volta,turing}. Each architecture has its microarchitecture and hardware characteristics. However, The percentage of undisclosed characteristics beyond what GPU vendors have documented is small. Hence, researches have proposed different micro-benchmarks written in programming languages, such as CUDA~\cite{Professional_CUDA_C} or OpenCL~\cite{opencl} to understand the hidden characteristics of the hardware for almost every GPU generations/architectures~\cite{moshovos,dissecting_gpu_mem,AMD_microbenchmarks,OpenCL_microbenchmarks,Intel_GPU_microbenchmarks}. Similarly, there are several works to develop assembly tool-chains that can provide direct access to the hardware using real machine-dependent opcodes~\cite{assemblers_fermi,assemblers_maxwell,bare_metal_perf_tuning,decuda}. These tool-chains usually provide more accurate results as they rely on low-level assembly language compared to the micro-benchmarks written in a relatively high-level language such as CUDA but they are not portable across different generations of GPUs.  

With each release of a new generation, a new version of the CUDA (\textit{nvcc}) compiler~\cite{nvcc} is usually released. NVIDIA has been constantly improving the CUDA compiler in terms of the techniques used to optimize the code. One type of code optimizations are machine dependent optimization which is done after the target code has been generated and when the code is transformed according to the target machine architecture. These optimizations affect the execution of individual instructions found in the ISA.

In this paper, we propose a low overhead and portable analysis to demystify the latency of different instructions executing in the pipeline and the different memory hierarchies found in various NVIDIA GPUs. 
We used parallel thread execution (PTX)~\cite{ptx} to perform our analysis. PTX is a pseudo-assembly language used in NVIDIA's CUDA programming environment. Alternatively, PTX can be described as a low-level parallel thread execution virtual machine which provides a stable programming model and instruction set for general-purpose parallel programming. PTX provides machine-independent ISA, thus the code is portable across different CUDA runtimes and GPUs. 
Using an assembly-like language such as PTX allows us to control the exact sequence of instructions executing in the pipeline and the type of accessed memory with low overhead. Since the compiler optimizations affect the instructions, we show the effect of the CUDA compiler optimizations on the execution of all instructions.

To the best of our knowledge, we are the first to provide an exhaustive analysis of different GPU instructions latencies. Moreover, no prior work discusses the effect of compiler optimizations on every single instruction executing in the pipeline. For this reason, we believe that this work is important, especially with the aggressive emergence of various technologies that rely on GPUs. There are multiple reasons why this characterization is important. \textbf{First}, it can give programmers more concrete understanding of the underlying hardware. Knowing the underlying microarchitecture would help GPU developers optimizing their applications' performance. Since the execution time of each kernel determines the application's overall performance, therefore, the programmer needs to be concerned with the execution time of each single instruction when writing high-performance code. Hence, it is critical to utilize hardware resources efficiently to achieve high performance. \textbf{Second}, GPUs software modeling frameworks, and cycle-accurate simulators~\cite{GPGPU-Sim,PPT_GPU,paraprox} depend on published instructions latencies in order to have an accurate model. Volkov~\cite{microbenchmark_to_study_GPU_perf} argued that inaccurate arithmetic instruction latencies, which are small but may accumulate to large numbers, can have a high impact on the accuracy of estimating the performance by these models. Since there exists no work in the literature that provides an in-depth GPU instruction latencies characterization, researchers had to collect the latencies along with other specifications from less academic sources such as graphics card databases and online reviews, especially for newer generations such as, Pascal~\cite{pascal} and Volta~\cite{volta}. \textbf{Third}, the effect of CUDA compiler optimizations on the instructions can guide GPU architects and code developers to choose what type of optimizations is needed and when.

Following are our \textbf{contributions}:
\begin{itemize}
    \item We provide a low overhead (really minimal) and portable method to estimate instructions latencies of modern GPUs. In addition, we show the overhead of accessing each memory hierarchy in these GPUs. 
    \item We show the effect of high-level optimizations found in the CUDA (\textit{nvcc}) compiler on different instructions.
    \item We provide an exhaustive comparison of all the instructions found in the PTX ISA.
    \item We run our evaluation on seven different high-end NVIDIA GPUs from five different GPU generations including the recently released Turing architecture
\end{itemize}
The rest of this paper is organized as follows: Section~\ref{sec:related} discusses relevant related work; whereas, Section~\ref{sec:background} explains the general architecture of NVIDIA GPUs; Section~\ref{sec:methodology} shows our methodology; while Section~\ref{sec:evaluation} shows the results; and finally, Section~\ref{sec:conc} concludes the paper.
\section{Related Work}
\label {sec:related}
\begin{figure}[t!]
      \centering
      \includegraphics[width=\linewidth]{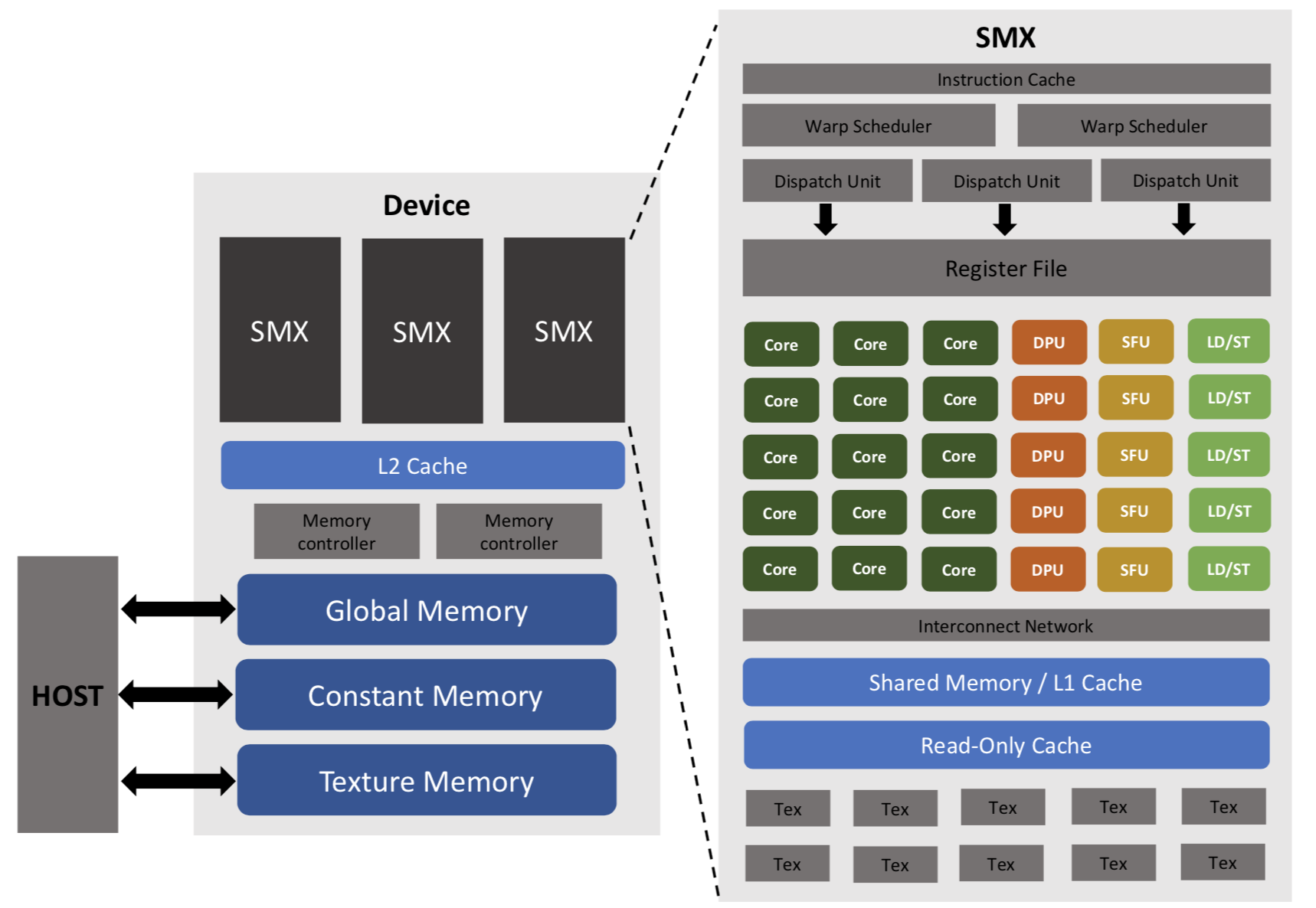}
\vspace{-3ex}
      \caption{Typical NVIDIA GPU architecture. The number of SMXs and the computation resources inside varies with the generation and the computational capabilities of the GPU.}
      \label{fig:gpu_arch}
\vspace{-3ex}
\end{figure}

Studying the hardware microarchitecture to undisclosed its hidden characteristics has been an active area of research for many years. Several micro-benchmarks were designed with the aim of dissecting the underlying CPU or GPU architecture. Furthermore, various studies have looked into tuning the application's source code to achieve better performance~\cite{implementation_dgemm_fermi,optimizing_sgemm_fermi_kepler,optimizing_stencil_kepler} but this task is tedious and requires a deep understanding of the underlying architecture. Hence, simulators, profilers, and optimization tools~\cite{GPGPU-Sim,PPT_GPU_memsys,GT-Pin,sassi,CUDAAdvisor,nvprof} were introduced to aid the architecture design space exploration. In this section, we discuss some of the related work in these areas in more details.

\textbf{Micro-benchmarks:} Wong~\etal~\cite{moshovos} have used micro-benchmarking to measure the latencies of some instructions and the characteristics of TLB and caches of an early NVIDIA Tesla generation GPU, (\textit{GeForce GT200}) . The work in~\cite{tune_dense_linear_algebra} measured the GPU kernel start-up costs, and arithmetic throughput to optimize dense linear algebra on (\textit{GeForce 8800GTX}) GPU which was released in 2006. In~\cite{dissecting_gpu_mem} the authors investigated the memory hierarchy of three different NVIDIA GPUs generations targeting their caches mechanism and latencies. Jia~\etal~\cite{dissecting_volta} studied the microarchitecture details of NVIDIA Volta (\textit{Tesla V100}) GPU architecture through micro-benchmarks and instruction set disassembly. 
The authors of~\cite{data_placement} used four different NVIDIA GPU generations to study the relevance of data placement optimizations of different GPU memories.

In summary, our work has much lower overhead compared to micro-benchmarking approaches similar to the assembly tool-chains~\cite{assemblers_fermi,assemblers_maxwell,bare_metal_perf_tuning,decuda} introduced in Section~\ref{sec:intro}. Hence, the results are more accurate. In addition, the code runs across different NVIDIA GPUs generations without sacrificing the ease of use nor the accuracy. On the other hand, micro-benchmarks, written in CUDA need to be designed specifically for each architecture and need to be updated manually with the emerge of a newer GPU generation.

\textbf{Compiler optimizations:} Chakrabarti~\etal~\cite{CUDA_optimization} described the effect of some CUDA compiler optimizations on computations written in CUDA running on GPUs.  In~\cite{autotuning_nvcc_param} the authors applied auto-tuning techniques to CUDA compiler parameters using the openTuner~\cite{OpenTuner} framework and compared the optimizations achieved by auto-tuning with the high-level optimization levels (\textit{-O0,-O1,-O2, and -O3}) found in the compiler. Yang~\etal~\cite{GPGPU_compiler_for_memory_opt} proposed an optimized GPU compiler framework which focuses on optimizing the memory usage of the application. They tested their framework on old NVIDIA GPUs (\textit{GeForce GTX8800}) and (\textit{GeForce GTX280}).

In summary, we follow the same line of research but we focus on the effect of the high-level optimization found in the CUDA compiler on individual instructions executing in the pipeline and on the access overhead of different memories found in modern GPUs.

\section{NVIDIA GPU Architecture Overview}
\label{sec:background}

A normal heterogeneous compute node nowadays consist of multicore CPU sockets connected to one or more GPUs. A GPU is currently not a standalone platform but rather a co-processor hosted by a CPU. Figure~\ref{fig:gpu_arch} shows a typical GPU architecture. The host (CPU) is connected to a PCIe bus to which the GPU board is also connected to. This means that the CPU sees the GPU as a PCIe device, thus it can access specific areas of the device memory to allocate and transfer data to. This includes global, constant, and texture memories in CUDA terminology.

The GPU architecture is built around an array of Steaming Multiprocessors (SMX) each can be seen as a standalone processor that can manage thousands of concurrent threads in single instruction multiple threads (SIMT) approach. Each SMX has a number of CUDA cores that has fully pipelined integer Arithmetic Units (ALUs) and floating-point units (FPU32) while being capable of executing one 32 bit integer or floating point operation per cycle. It also includes Double-Precision Units (DPU) for 64-bit computations, Special Function Units (SFU) that executes intrinsic instructions, Load and Store units (LD/ST) for calculations of source and destination memory addresses. In addition to the computational resources, each SMX is coupled with a certain number of warp schedulers, instruction dispatch units, instruction buffer(s) along with texture and shared memory units.

\textbf{CUDA Memory Model.} Both GPUs and CPUs use similar principals in memory hierarchy design. The key difference is that in GPUs, the memory hierarchy is more exposed and this gives the programmer more explicit control over its behavior. Each memory space in the GPU has a different scope, lifetime, and caching behavior. Global, constant, and texture memories reside in the device memory, thus they have high access latencies and their content have the same lifetime as the running application. On the other hand, shared memory contents have the same lifetime as a thread block in a CUDA kernel with much lower access latency.

\textit{Global Memory} is the largest, and most commonly used memory, which is accessed by all threads from an SMX. The global memory contents are cached in two levels. There is a small L1 cache per SMX and a L2 cache shared by all SMXs.

\textit{Local Memory} is for register spilling. Any kernel variable that do not fit in registers, will be spilled to the local memory. Local memory data is cached the same way as global memory.

\textit{Constant Memory} is for data that will not change with the kernel execution and is cached in a dedicated per SMX read-only cache. The logical constant space can be allocated on the device memory is 64KB for different compute capabilities.

\textit{Texture Memory} is originally designed for traditional graphics applications but now it can be used as a read-only memory that can improve performance and reduce memory traffic when reads have certain access patterns. It is a dedicated per SMX read-only memory like constant memory.

\textit{Shared Memory} is a programmable memory used in the communication among threads in a block. It is an on-chip per SMX memory with high bandwidth and low access latency.

In Kepler, Volta, and Turing, the L1 data cache and the shared memory  physically share the same space, while on Maxwell and Pascal the L1 data cache is separated from the shared memory and combined with texture cache.
\section{Methodology}
\label{sec:methodology}

\begin{figure}
      \centering
      \includegraphics[width=\linewidth]{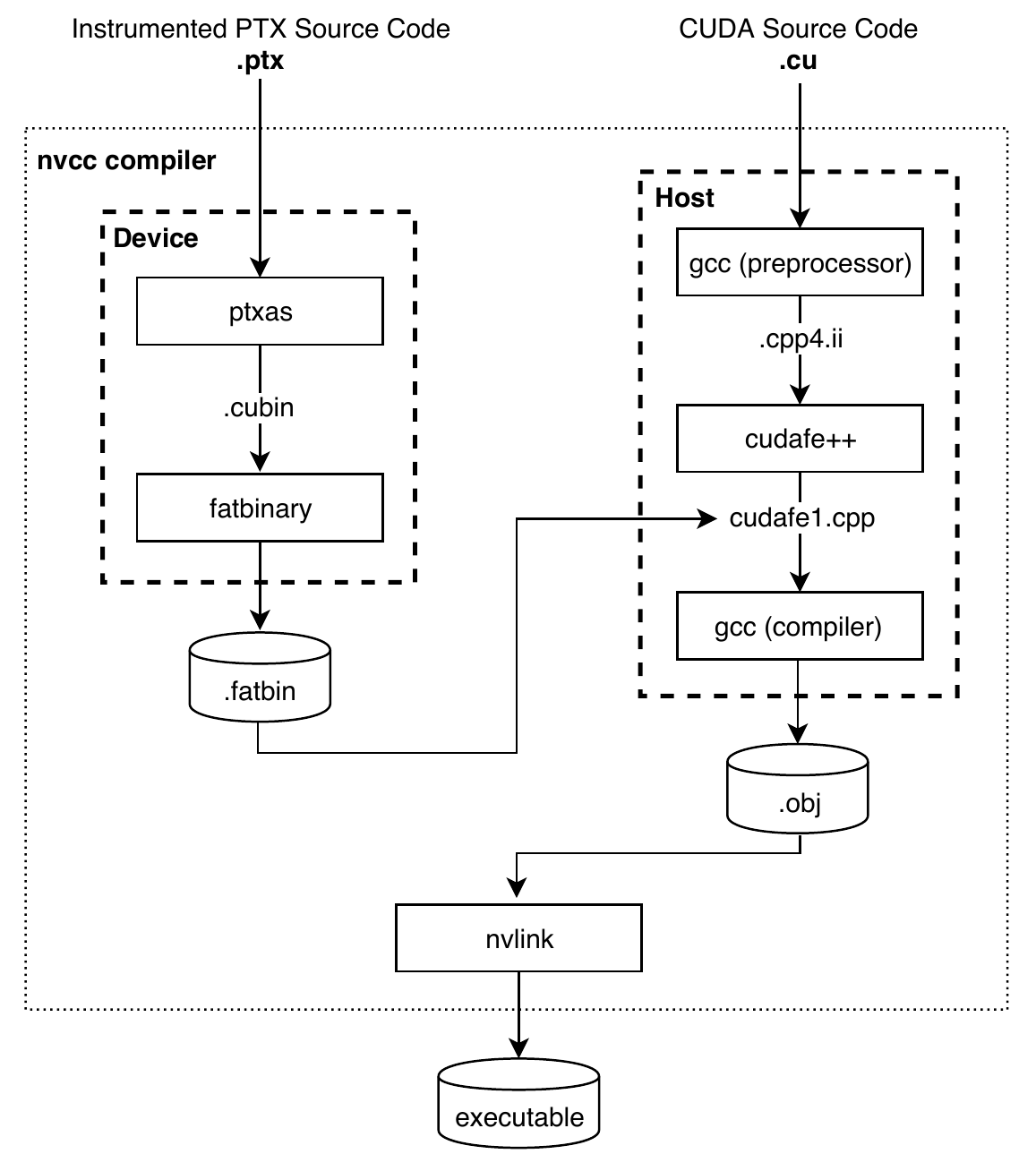}
\vspace{-3ex}
      \caption{Overview of the compilation workflow. }
      \label{fig:workflow}
\vspace{-3ex}
\end{figure}

\definecolor{backcolour}{rgb}{0.95,0.95,0.92}
\definecolor{codepurple}{rgb}{0.58,0,0.82}
\definecolor{codegreen}{rgb}{0,0.6,0}
\definecolor{codegray}{rgb}{0.5,0.5,0.5}

\lstdefinelanguage{PTX}
{
  morekeywords={ld, st, mov, add, sub},
  ndkeywords={global, clock, param, ret}
}

\begin{figure*}
    \centering
    \begin{minipage}{.5\textwidth}
        \centering
           \begin{lstlisting}[language=PTX,
                      numbers=left, 
                      xleftmargin=0.3in, 
                      xrightmargin=0.3in, 
                      basicstyle=\scriptsize,
                      frame=single, 
                      backgroundcolor=\color{backcolour},
                      numberstyle=\scriptsize\color{codegray},
                      keywordstyle=\color{codepurple},
                      ndkeywordstyle=\color{blue}
                     ]
.visible .entry Add(
    .param .u64 Add_param_0,
    .param .u64 Add_param_1,
    .param .u64 Add_param_2
){
    .reg .b32   %r<7>;
    .reg .b64   %rd<4>;
    
    ld.param.u64    %rd1, [Add_param_0];
    ld.param.u64    %rd2, [Add_param_1];
    ld.param.u64    %rd3, [Add_param_2];

    ld.global.u32   %r4, [%rd1];
    ld.global.u32   %r5, [%rd2];

    mov.u32         %r1, %clock;
    add.u32         %r6, %r4, %r5;
    mov.u32         %r2, %clock;
    sub.s32         %r3, %r2, %r1;

    st.global.u32   [%rd3], %r3;
    
    ret;
}
\end{lstlisting} 
\vspace{-1ex}
\caption{Computing unsigned add instruction latency.}
\vspace{-2ex}
        \label{fig:PTX_ALU}
    \end{minipage}%
    \begin{minipage}{0.5\textwidth}
        \centering
        \begin{lstlisting}[language=PTX,
                    numbers=left, 
                    xleftmargin=0.3in, 
                    xrightmargin=0.3in, 
                    basicstyle=\scriptsize,
                    frame=single, 
                    backgroundcolor=\color{backcolour},
                    numberstyle=\scriptsize\color{codegray},
                    keywordstyle=\color{codepurple},
                    ndkeywordstyle=\color{blue}
                 ]
.visible .entry globalMem(
    .param .u64 globalMem_param_0,
    .param .u64 globalMem_param_1,
){
    .reg .b32   %r<5>;
    .reg .b64 	%rd<3>;

    ld.param.u64    %rd1, [globalMem_param_0];
    ld.param.u64    %rd2, [globalMem_param_1];

    mov.u32         %r1, %clock;
    ld.global.u32   %r4, [%rd1 + 4];
    mov.u32         %r2, %clock;
    sub.s32         %r3, %r2, %r1;

    mov.u32         %r1, %clock;
    ld.global.u32   %r5, [%rd1 + 8];
    mov.u32         %r2, %clock;
    sub.s32         %r4, %r2, %r1;

    st.global.u32   [%rd2], %r3;
    st.global.u32   [%rd2 + 4], %r4;
    ret;
}
\end{lstlisting}
\vspace{-1ex}
        \caption{Computing device and L1/L2 memories access latencies.}
\vspace{-2ex}
         \label{fig:PTX_mem}
    \end{minipage}
\vspace{-1ex}
\end{figure*}

In this section, we describe our approach. The instructions timing model (Section~\ref{subsec:timing_model}) is written in PTX~\cite{ptx}. PTX is a virtual assembly ISA that is forward-compatible to all NVIDIA architectures and generations. PTX allows us to have control over the exact sequence of low-level instructions executing without any loop or any other CUDA overhead. Since PTX is considered as a virtual ISA, it gets translated to another machine assembly ISA that gets executed on the GPU known as Source And Assembly (SASS). SASS is only forward-compatible within the same major family (\textit{Fermi, Kepler,}~\etc). SASS is not open-sourced and its instructions and characteristics are not well-documented and require CUDA Binary Utilities~\cite{binary_utilities} and reverse-engineering tools to disassemble. 

Figure~\ref{fig:workflow} shows the compilation workflow which relies on CUDA nvcc compiler~\cite{nvcc}. The instrumented PTX source code which contains the instructions timing model is first compiled with the PTX optimizing assembler (\textit{ptxas}) to produce a device CUDA binary file (\textit{.cubin}) which is in SASS. This binary file is then placed in a fatbinary (\textit{.fatbin}) which gets embedded in the host input source code file. The embedded fatbinary is inspected by the CUDA runtime system whenever the device code is launched by the host program to obtain an appropriate fatbinary image for the required GPU family. A single object file (\textit{.obj}) containing the host and the device source code is then generated and linked to produce an executable file. 

\subsection{Timing Model}
\label{subsec:timing_model}
To determine each operation latency 
We read the \textit{clock} register before and after the execution of the instruction. The \textit{clock()} function provides a per-multiprocessor counter that is incremented every clock cycle. Sampling this counter at the beginning and at the end of the operation and taking the difference between the two samples gives us the exact number of cycles this operation takes to finish execution. Reading the \textit{clock} register in PTX is translated to register read followed by a dependent operation in SASS. Thus, we calculate the \textit{clock} function overhead in order to subtract it later from the time obtained for each operation. 

Figure~\ref{fig:PTX_ALU} shows an example for obtaining the latency of unsigned integer add instruction. Two scalar variables are passed to the kernel and loaded into two registers (line 8 to 14). In line 16, we read the clock register followed by an \textit{add} instruction and reading the clock register again. The results we have from subtracting the two values of the clock register is then subtracted from the clock overhead (Section~\ref{subsubsec:clock_overhead}) to obtain the exact number of cycles the hardware took to execute the instruction, in this case the unsigned add instruction.

Figure~\ref{fig:PTX_mem} shows an example for obtaining the latency for accessing the device (global memory) and the cache memories of the GPU. The exact same approach of figure~\ref{fig:PTX_ALU} is used but this time we pass a vector to the kernel so that we can calculate the caches hit latency. In line 12, the \textit{load} instruction will go all the way to fetch the block since it is a cold cache. This will give us the access time of the global memory. We leverage the option provided by the CUDA compiler to tweak the application to disable or enable the L1 cache while compiling it. We compile the application two times, first using L1 and L2 caches thus the block is fetched from the global memory when loading a block (line 12) and put it in L2 and L1 caches. Hence, when loading a value from the same block again (line 17) it is a hit in the L1 cache and this gives us the hit latency of the L1 cache. We do the same thing while disabling the L1 cache and forcing the application to use the L2 cache only. This would get us the hit latency of the L2 cache. We make sure not to read the exact value again in line 17 but rather a new value from the same cache block so that the compiler does not optimize it and change it to a regular \textit{mov} instruction. 

Since we only care about the individual instructions latencies and not the overall throughput, the kernels were executed with only one thread per warp. To show the effect of the compiler optimizations, we compile the code using the high-level optimization flags found in the CUDA compiler (\textit{-O0, -O1, -O2, -O3)}. To make sure that the hardware executes the instructions without interference and/or truncation of instructions by the compiler in the \textit{-O3} level, we perform a dependent dummy operation on the output of the instruction. Sometimes, the compiler reorders the kernel instructions when translating from PTX to SASS. This can move the instructions out of the clock timing block. Therefore, we added memory and thread barriers just to make sure that the code gets translated verbatim as is and the instruction(s) is (are) inside the clock timing block. 
\section{Evaluation}
\label{sec:evaluation}

\newcolumntype{P}[1]{>{\centering\arraybackslash}p{#1}}
\renewcommand{\arraystretch}{1.2}
\begin{table}[t!]\tiny
\centering
  \caption{Target GPUs Configurations.}
\vspace{-3ex}
    \begin{tabular}{|P{1.3cm}||P{1.07cm}|P{1cm}|P{1.07cm}|P{0.9cm}|P{0.9cm}|}
  \hline
  \textbf{Configuration}  &\textbf{K40m}   &\textbf{TITAN}   &\textbf{P100}   &\textbf{V100} &\textbf{TITAN}  \\
  &&\textbf{X}&&&\textbf{RTX}\\
  \hline
  &\multicolumn{5}{c|}{\textbf{Graphics Processor}} \\
  \hline
   Architecture         & Kepler  & Maxwell  & Pascal   & Volta    & Turing \\
  \hline
   Chip                 & GK110B  & GM200    & GP100    & GV100    & TU102 \\
  \hline
   Compute Capability   & 3.5     & 5.2      & 6.0      & 7.0      & 7.5\\
  \hline
  &\multicolumn{5}{c|}{\textbf{Clock Speeds}} \\
  \hline
   GPU Clock     & 745 MHz   & 1000 MHz  & 1190 MHz   & 1246 MHz   & 1350 MHz\\
  \hline
   Memory Clock  & 1502 MHz  & 1753 MHz  & 715 MHz    & 876 MHz    & 1750 MHz \\
  \hline
  &\multicolumn{5}{c|}{\textbf{Memories}} \\
  \hline
   Memory Size      & 12 GB       & 12 GB      &  16 GB       &  16 GB   & 24 GB\\
  \hline
   Memory Type      & GDDR5       & GDDR5      &  HBM2        &  HBM2    & GDDR6\\
  \hline
   Memory Bus       & 384 bit     & 384 bit    &  4096 bit    & 4096 bit & 384 bit \\
  \hline
   Memory Bandwidth & 288.4 GB/s  & 336.6 GB/s &  732.2 GB/s  & 897.0 GB/s & 672.0 GB/s \\
  \hline
   L1 Size          & 16 KB       & 24 KB      &  48 KB       & 128 KB   & 64 KB \\
  \hline
   L2 Size          & 1536 KB     & 3 MB       &  4 MB        & 6 MB     & 6 MB \\
  \hline
  &\multicolumn{5}{c|}{\textbf{Theoretical Performance (TFLOPS)}} \\
  \hline
   FP16 (half)     &  NA   & NA    & 19.05    &  28.26    & 32.62 \\
  \hline
   FP32 (float)    &  5.046   & 6.691   &  9.526   &  15.7    & 16.31 \\
  \hline
   FP64 (double)   &  1.682   & 0.2061  &  4.763   & 7.8  & 0.5098 \\
  \hline
   Texture Rate (GTexel/s)  &  210.2   &  209.1   &  297.7   & 441.6   & 509.8  \\
  \hline
  &\multicolumn{5}{c|}{\textbf{SMX Level}} \\
  \hline
  \# \{Cores, SMX\}      &\{2880, 15\}  &\{3072, 24\}  &\{3584, 56\}  &\{5120, 80\}   &\{4608, 73\} \\
  \hline
  \# \{SP, DP, SFU\}    &\{192, 64, 32\}    &\{128, 4, 32\}    &\{64, 32, 16\}   &\{64, 32, 4\}   &\{64, 2, 4\}\\
  \hline
  \# LD/ST         & 32           & 32           &  16         &16      & 16\\
  \hline
  \end{tabular}
  \label{table:gpus_configs}
\vspace{-2ex}
\end{table}

\begin{figure}[t!]
	\centering \includegraphics[width=\linewidth]{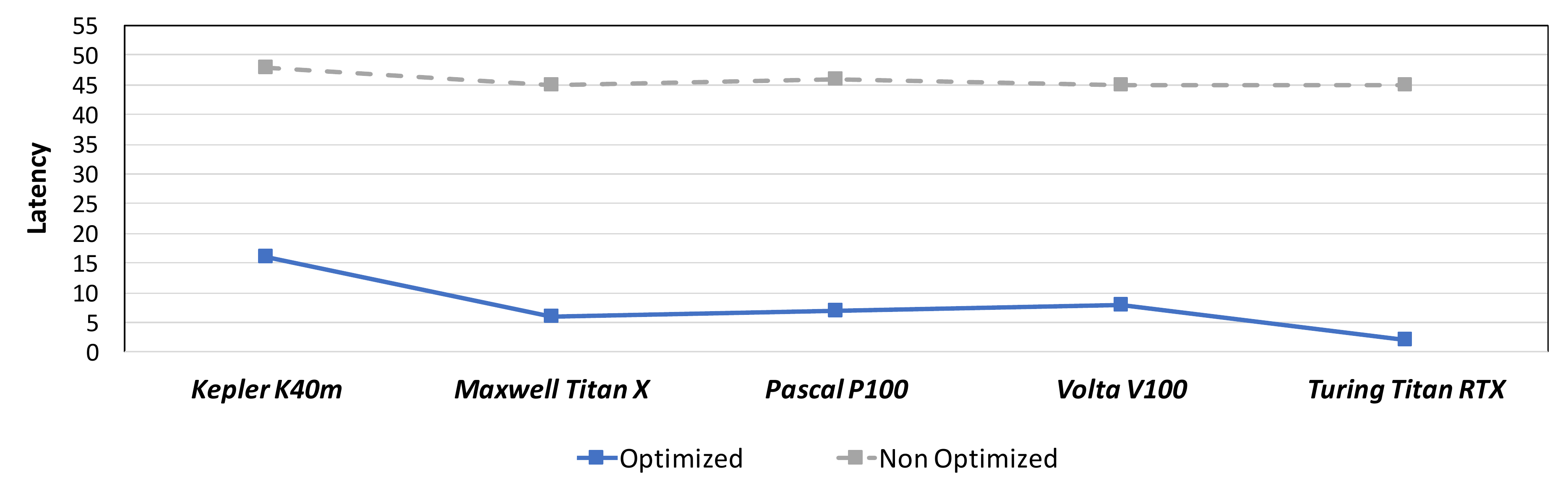}
\vspace{-2ex}
	\caption{Clock Overhead}
   \label{fig:clk_overhead}
\vspace{-2ex}
\end{figure}

\begin{figure*}
\centering
\subfigure[Global Memory \& L1 / L2 Caches Access Latencies]{%
\label{fig:global_mem}%
\includegraphics[width=3.4in]{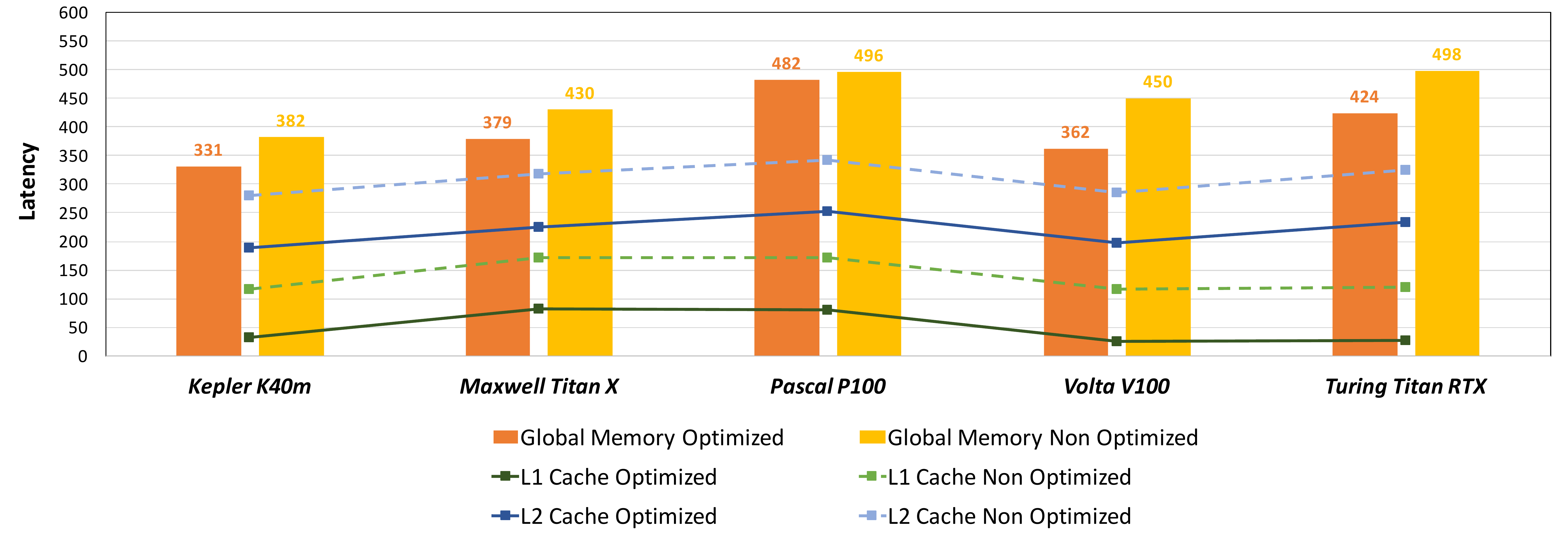}}%
\qquad
\subfigure[Texture Memories \& Texture Cache Access Latencies]{%
\label{fig:text_mem}%
\includegraphics[width=3.4in]{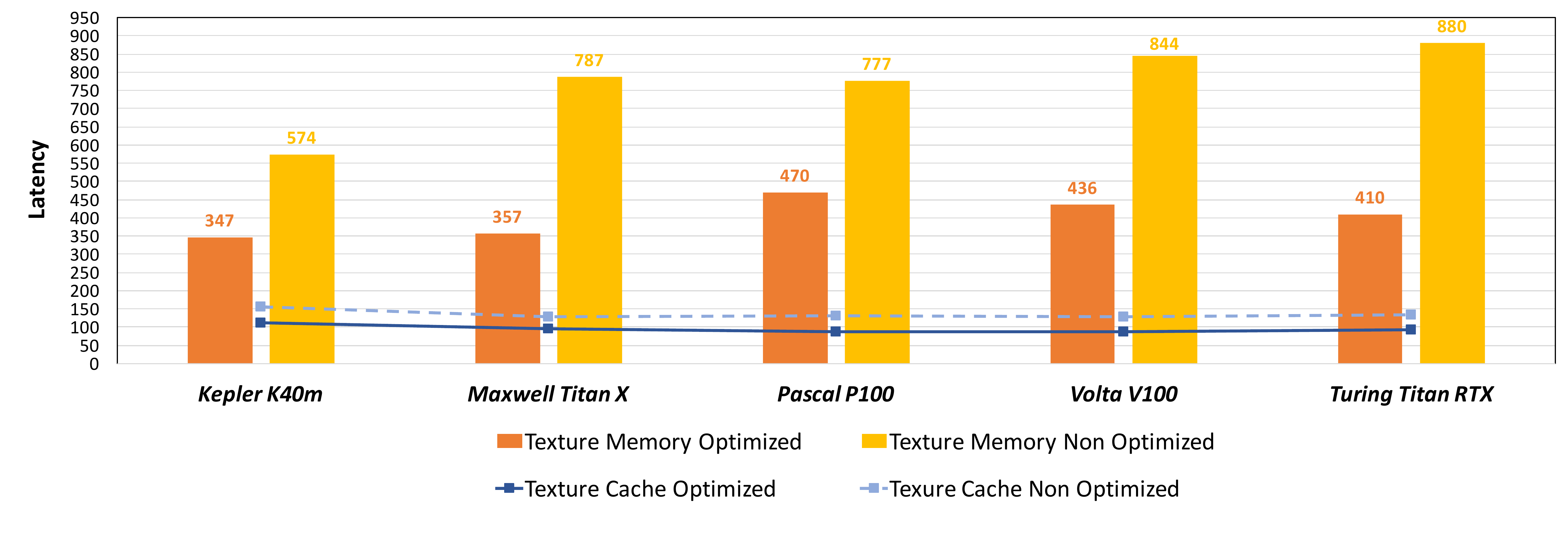}}%
\vspace{-1ex}
\caption{Different Memory Units Access Overhead}
\vspace{-2ex}
\end{figure*}

In this section, we illustrate the differences in characteristics between the various GPUs tested in this paper. 

\subsection{Target GPUs}
\label{subsec:evaluation_setup}
We run our evaluation on seven different high-end GPUs from five different generations (\textit{Kepler~\cite{kepler}, Maxwell~\cite{maxwell}, Pascal~\cite{pascal}, Volta~\cite{volta}, and Turing~\cite{turing}}). Table~\ref{table:gpus_configs} depicts the differences in configurations and theoretical performance between the main five GPUs evaluated. The additional two GPUs are K80c and TITAN V which are from the same generation as K40m (\textit{Kepler}) and V100 (\textit{Volta}) respectively. We used these two additional GPUs to verify if the results change within the same architecture by the particular model of the GPU. 
Most of the configuration parameters are collected from NVIDIA's white papers and non-academic sources such as graphics cards databases and online reviews.

\subsection{Evaluation Results}
\label{subsec:evaluation_results}

We divide the instructions into two categories; instructions that use the computational units in the GPU, termed \textit{ALU Instructions} and data movement instructions that use the different memories in a GPU, termed \textit{Memory Instructions}. We used CUDA version 9.0~\cite{cuda_9.0} for all our assessment except for the TITAN RTX GPU which supports only CUDA version 10.0~\cite{cuda_10.0}. We run the code with the different optimization levels in the {\em nvcc} compiler. Due to space, we only show the results of (\textit{-O3}) and (\textit{-O0}) which we denote as \textit{Optimized} and \textit{Non Optimized} in the results. The other two optimization levels have almost the same results as the \textit{Optimized} results provided here. In order to see whether different CUDA versions will have an effect on our results, we used CUDA version 10.0~\cite{cuda_10.0} on Volta GPUs and compared the results with the results we got while using CUDA version 9.0.

\subsubsection{\textbf{Clock Overhead}}
\label{subsubsec:clock_overhead}
We first calculate the \textit{clock} function overhead. Reading the \textit{clock} register is processed on the hardware by a \textit{move} instruction followed by a dependant operation. Figure~\ref{fig:clk_overhead} shows the difference in \textit{clock} overhead between the different GPU architecture. It also shows the effect of the optimization levels on reading the \textit{clock} register.

\subsubsection{\textbf{ALU Instructions}}
\label{subsubsec:alu_instructions}
Table~\ref{table:alu_results} in the Appendix shows the instruction overhead latencies for different NVIDIA GPUs. We run all the instructions found in the latest PTX version released,  6.4~\cite{ptx}. We divide the instructions into eight different categories and group the instructions whose latencies are the same together. We group the GPUs that are from the same generations together and if both have the same results for the same instruction we write only one number in that entry in the table.

We noticed that some instructions such as \textit{div} and \textit{rem} have different results when operating on signed an unsigned numbers. We denote that by \{s\} and \{u\} respectively. In addition, all the instructions have the same output when operating on different data values except \textit{div} instruction which have different output depending on the data values. This is mainly because it gets optimized and changed by the compiler into shift operations when the divisor is a power of two. Thus, We denote that by (regular), (irregular), and (average) which state if the divisor is a power of two, not a power of two, and the average between the two cases respectively. 

\textit{Half Precision (FP16)} instructions were first supported with the releasing of Pascal architecture GPUs. This was an artifact of the prevalence of approximate computing especially for machine/deep learning acceleration~\cite{AxBench,exploiting_half_precision}. Although both P100, V100, and RTX have the same numbers for the (FP16) instructions, the Turing architecture has higher theoretical performance than the other two as shown in table~\ref{table:gpus_configs}.

\textit{Multi Precision} instructions combine the use of different numerical precision in a computational method. It offers significant computational speedups by performing operations in half-precision format, while storing minimal information in single-precision to retain as much information as possible. This is used in critical parts of neural networks~\cite{mixed_precision_training}. If the architecture does not support half-precision it uses single-precision only. Table~\ref{table:alu_results} in the Appendix shows that Volta and Turing architectures have the best results across all the other generations.

The ALU instruction latencies have significantly decreased from the Kepler architecture to successor Maxwell, except for the \textit{div} instruction. Maxwell and Pascal architectures have very close results, however, Maxwell has lower double precision performance. Turing architecture has the best results in the \textit{Integer Arithmetic Instructions} but it exhibits very high latencies in double precision operations. In half and multi precision instructions, Turing and Volta have almost the same results but Turing has higher throughput. Hence, approximate computing applications can experience very good performance on such architecture. On the other hand, Volta GPUs are unbeatable in single and double precision floating point performance.

We run the same evaluation using CUDA compiler version 10.0 on Volta architecture GPUs to see whether different CUDA versions will affect the optimizations done on the instructions. Table~\ref{table:volta_cuda_10} shows the instructions which experienced differences in latencies between the two version. From these results, we can confidently conclude that CUDA compiler version 10.0 has lower latencies, thus better optimizations.

\renewcommand{\arraystretch}{1.3}
\begin{table}[b!]
  \centering
  \caption{Optimizations effect of different CUDA Compiler Versions on VOLTA (\textit{TITANV / V100}) GPUs}
\vspace{-2ex}
  \begin{tabular}{|c||c|c|}
  \hline
  \textbf{Instruction} & \textbf{CUDA Version 9.0} & \textbf{CUDA Version 10.0}  \\
  \hline
  \multicolumn{3}{|c|}{\textbf{Floating Single Precision Instructions}} \\
 \hline
   div (regular)    & 123  & 116 \\
  \hline
   div (irregular)  & 280   & 266  \\
  \hline
   div (average)    & 201   & 191  \\
  \hline
  \multicolumn{3}{|c|}{\textbf{Double Precision Instructions}} \\
  \hline
   div   & 159 & 135 \\
  \hline
  \multicolumn{3}{|c|}{\textbf{Integer Intrinsic Instructions}} \\
  \hline
   mul64hi()          & 123 & 85 \\
  \hline
   popc()             & 15 & 5 \\
  \hline
   bfind() / bbrev()  & 15 & 5 \\
  \hline
  \end{tabular}
   \label{table:volta_cuda_10}
\end{table}

\renewcommand{\arraystretch}{1.3}
\begin{table}[b!]
  \centering
  \caption{Shared \& Constant Memories Access Latency }
\vspace{-2ex}
  \begin{tabular}{|c||c|c|c|c|c|}
  \hline
  \textbf{Memory Unit} & \textbf{K40m} & \textbf{TITAN X} & \textbf{P100}  & \textbf{V100} & \textbf{RTX}\\
  \hline
   \multirow{5}{*}{\textbf{Shared Memory}}
   &\multicolumn{5}{c|}{\textit{\textbf{Optimized}}}\\
   \cline{2-6}
   
   & 26   & 24    & 25   & 18    & 21  \\
   
   \cline{2-6}
   &\multicolumn{5}{c|}{\textit{\textbf{Non Optimized}}}\\
   \cline{2-6}
   
   & 55   & 53    & 54   & 49   &  37 \\
  \hline

  \multirow{5}{*}{\textbf{Constant Memory}}
  &\multicolumn{5}{c|}{\textit{\textbf{Optimized}}}\\ 
  \cline{2-6}
  
  & 16   & 20    & 12   & 8    &   8  \\
  
  \cline{2-6}
  &\multicolumn{5}{c|}{\textit{\textbf{Non Optimized}}}\\
  \cline{2-6}
  
  & 80   & 145   & 71   & 70   &  71   \\
  \hline
  \end{tabular}
   \label{table:shared_constant_mem}
\vspace{-1.5ex}
\end{table}

\subsubsection{\textbf{Memory Instructions}}
\label{subsubsec:memory_instructions}
The global and texture memories access overhead is shown in figures~\ref{fig:global_mem} and~\ref{fig:text_mem} respectively. The figures show that the difference between the access latencies is nearly unnoticeable. NVIDIA focused more on increasing the bandwidth of the main memory, the memory interconnect (bus), and the texture rate rather than the access latency since the latency is going to be tolerated by thread level parallelism. Despite the fact that Kepler has nearly the same access latency for the global and the texture memories as Volta, Volta has more than double the memory bandwidth and the texture rate compared to Kepler as shown in table~\ref{table:gpus_configs}. The figures also show that the non-optimized version of the texture memory is nearly double that of the optimized version which is not the case in the global memory.

The L1 data cache have higher access latency on Maxwell and Pascal which comply with the fact that they share the same physical space with the texture cache. On the hand, combining the L1 data cache and the shared memory on Kepler, Volta, and Turing significantly reduces the cache hit latency and improves the bandwidth. The L2 data cache experience very high access latency due to bank conflicts which sometimes leads to memory divergence and forces many of these requests to queue up for long periods of time~\cite{mutlu_memory_queue}. However, with the release of new architectures, the size of the L2 caches usually increases and this improves the bank conflicts.

Table~\ref{table:shared_constant_mem} shows the shared and constant memories access latency. They both have very low latency compared to other memories. The constant cache memory gets optimized by the CUDA compiler to be almost as a register to register latency.

\section{Conclusion}
\label {sec:conc}
In this paper, we benchmark the undisclosed instructions latencies and different memory access overhead(s) of various generations of the NVIDIA GPGPUs. We also show the effect of the different optimization levels in the CUDA (\textit{nvcc}) compiler on the individual ptx  instructions. We run our evaluation on seven different NVIDIA GPUs from five different GPU architectures. Our results show that the instructions latencies have mostly decreased from Kepler to Turing. These results should help architects and software developers in optimizing both the hardware and the software and to understand the impact and sensitivity of applications on the various GPU generations.



\section*{Appendix I: The Latency of the Various ALU Instructions}
This Appendix contains an enumeration of the latencies of the various ALU instructions for the various GPUs.
\renewcommand{\arraystretch}{1.3}
\begin{table*}
\centering
  \caption{The Latency of the Various ALU Instructions.}
\vspace{-2ex}
  \begin{tabular}{|P{2.7cm}||P{1.33cm}|P{0.9cm}|P{0.9cm}|P{1.2cm}|P{0.8cm}||P{1.38cm}|P{0.9cm}|P{0.9cm}|P{1.2cm}|P{0.8cm}|}
  \hline
   \multirow{4}{*}{\textbf{Instruction}} &\multicolumn{5}{c||}{\textbf{Optimized}} & \multicolumn{5}{c|}{\textbf{Non Optimized}} \\
  \cline{2-11}
  &\textbf{K40m / K80c} & \textbf{TITAN X} & \textbf{P100} & \textbf{TITAN V / V100} &\textbf{TITAN RTX} &\textbf{K40m / K80c}  & \textbf{TITAN X} & \textbf{P100} & \textbf{TITAN V / V100} &\textbf{TITAN RTX}\\
  \cline{2-11}
  &\multicolumn{10}{c|}{\textbf{(1) Integer Arithmetic Instructions}} \\
  \hline
   add / sub / min / max    & 9   & 6    & 6   & 4   & 4  & 16   & 15    & 15   & 15  & 15\\
  \hline
   mul / mad                & 9   & 13   & 13  & 4   & 4  & 16   & 87    & 85   & 15  & 15\\
  \hline
  \{s\} div (regular)       & 134 & 141  & 144 & 125 & 117  & 791  & 1020  & 1039 & 815 & 785\\
  \hline
  \{s\} div (irregular)     & 164 & 160  & 163 & 129 & 121  & 791  & 1020  & 1039 & 815 & 785\\
  \hline
  \{s\} div (average)       & 149 & 150  & 153 & 127 & 119  & 791  & 1020  & 1039 & 815 & 785\\
  \hline
  \{s\} rem                 & 132 & 141  & 144 & 125 & 114  & 751  & 955   & 1017 & 770 & 740\\
  \hline
  abs                       & 16  & 13   & 13  & 8   & 8  & 32   & 30    & 30   & 30  & 45\\
  \hline
  \{u\} div (regular)       & 123 & 127  & 130 & 120 & 112  & 608  & 856   & 851  & 619 & 589\\
  \hline
  \{u\} div (irregular)     & 140 & 146  & 149 & 125 & 116  & 608  & 856   & 851  & 619 & 589\\
  \hline
  \{u\} div (average)       & 131 & 136  & 139 & 122 & 114  & 608  & 856   & 851  & 619 & 589\\
  \hline
  \{u\} rem                 & 116 & 127  & 130 & 117 & 109  & 576  & 826   & 821  & 590 & 560\\
  \hline
  &\multicolumn{10}{c|}{\textbf{(2) Logic and Shift Instructions}} \\
 \hline
   and / or / not / xor     & 9   & 6  & 6   & 4  & 4  & 16  & 15  & 15  & 15  & 15\\
  \hline
   cnot                     & 18  & 6  & 12  & 8  & 8  & 48  & 45  & 45  & 45  & 45\\
  \hline
   shl/shr                  & 9   & 6  & 6   & 4  & 4  & 16  & 15  & 15  & 15  & 15\\
  \hline
  &\multicolumn{10}{c|}{\textbf{(3) Floating Single Precision Instructions}} \\
 \hline
   add / sub / min / max    & 9  & 6  & 6   & 4   & 4  & 16  & 15    & 15  & 15  & 15\\
  \hline
   mul / mad / fma          & 9  & 6  & 6   & 4   & 4  & 16  & 15    & 15   & 15  & 15\\
  \hline
   div (regular)        & 151 / 150  & 135  & 167 & 123 & 152   & 661 / 629 & 725  & 671  & 638 & 546\\
  \hline
   div (irregular)      & 686 / 479  & 765  & 649 & 280 & 303   & 661 / 629 & 725  & 671  & 638 & 546\\
  \hline
   div (average)        & 418 / 314   & 450  & 408 & 201 & 227  & 661 / 629 & 725  & 671  & 638 & 546\\
  \hline
  &\multicolumn{10}{c|}{\textbf{(4) Double Precision Instructions}} \\
 \hline
   add / sub / min / max   & 10   & 48  & 8   & 8   & 40  & 16  & 52 & 15   & 15  & 48\\
  \hline
   mul / mad / fma          & 10  & 48  & 8   & 8   & 40  & 16  & 52 & 15   & 15  & 54\\
  \hline
   div (average)          & 445 / 428 & 709  & 545 & 159 & 540  & 1588 / 1338 & 1821  & 1399 & 945 & 1202\\
  \hline
  &\multicolumn{10}{c|}{\textbf{(5) Half Precision Instructions}} \\
 \hline
   add / sub   & NA  & NA  & 6  & 6  & 6  & NA  & NA  & 15  & 15  & 15 \\
  \hline
   mul         & NA  & NA  & 6  & 6  & 6  & NA  & NA  & 15  & 15  & 15 \\
  \hline
   fma         & NA  & NA  & 6  & 6  & 6  & NA  & NA  & 15  & 15  & 15 \\
  \hline
  &\multicolumn{10}{c|}{\textbf{(6) Multi Precision Instructions}} \\
 \hline
   add.cc / addc / sub.cc    & 9   & 6  & 6   & 4   & 4  & 16  & 15  & 15  & 15  & 15   \\
  \hline
   subc                      & 18  & 12  & 12  & 8  & 8  & 32  & 30  & 30  & 30  & 30  \\
  \hline
   mad.cc/madc               & 9   & 13  & 13  & 4  & 4  & 16  & 87  & 85  & 15  & 15  \\
  \hline
   &\multicolumn{10}{c|}{\textbf{(7) Special Mathematical Instructions}} \\
 \hline
  rcp           & 377 / 298 & 347  & 266 & 60 & 92 & 459 / 429 & 534 & 395 & 316 & 315 \\
  \hline
  sqrt          & 432 / 352 & 360  & 282 & 60 & 96 & 465 / 431 & 540  & 399 & 330 & 330 \\
  \hline
  fast approximate sqrt & 49 & 47  & 35  & 31 & 31 & 304  & 285  & 540 & 270 & 270 \\
  \hline
  fast approximate rsqrt & 40 & 34  & 35  & 31 & 31 & 288  & 270  & 270 & 270 & 270 \\
  \hline
  fast approximate sin/cos  & 18 & 15  & 15  & 11 & 13 & 32  & 30  & 30  & 30  & 30 \\
  \hline
  fast approximate lg2  & 40 & 34  & 35  & 31 & 31 & 288  & 270  & 270 & 270 & 270 \\
  \hline
  fast approximate ex2  & 49  & 40  & 41  & 22 & 32 & 256  & 240  & 240 & 225 & 225 \\
  \hline
  copysign  & 21 & 20  & 20  & 8  & 7 & 80  & 75  & 75  & 75  & 75 \\
  \hline
     &\multicolumn{10}{c|}{\textbf{(8) Integer Intrinsic Instructions}} \\
 \hline
  mul24() / mad24()    & 22   & 21   & 21   & 12  & 12   & 48    & 118   & 116   & 75    & 75 \\
  \hline
  mulhi()              & 9    & 18   & 18   & 12  & 8    & 16    & 85    & 86    & 32    & 17 \\
  \hline
  mul64hi()            & 226  & 118  & 123  & 106 & 106  & 896   & 1419  & 1420  & 578   & 578 \\
  \hline
  sad()                & 9    & 6    & 6    & 4   & 4    & 16    & 15    & 15    & 15    & 15 \\
  \hline
  popc()               & 9    & 13   & 13   & 15  & 15   & 32    & 45    & 45    & 45    & 45 \\
  \hline
  clz()                & 20   & 19   & 18   & 5   & 21    & 32    & 30    & 30    & 30    & 30 \\
  \hline
  bfe() / bfi()        & 9    & 6    & 6    & 4   & 4    & 16    & 15    & 15    & 15    & 15 \\
  \hline
  bfind() / bbrev()    & 9    & 6    & 6    & 15  & 15    & 48    & 45    & 45    & 45    & 45 \\
  \hline
  \end{tabular}
  \label{table:alu_results}
\end{table*}

\end{document}